# CFOs Meet LLMs

John R. Graham[†] Campbell R. Harvey[†] Manish Jha[‡]

[†]Duke University and NBER [‡]Georgia State University

**Abstract**

Business sentiment is closely watched, but measurement is slow and costly: surveys reach a limited number of firms, arrive periodically, and take time to field and compile. We show that large language models hold the potential to address some of these shortcomings. We prompt an LLM to role-play as the CFO of a specific company on a specific date, for every public-company CFO who responded to the Duke–Federal Reserve CFO Survey from 2002- 2025. We find that LLM-generated forecasts reasonably match the cross-section of CFO expectations, with accuracy increasing with the amount of information provided to the LLM.

Version: November 15, 2026. The authors have no conflicts of interest to disclose. We thank Manuel Adelino, Alon Brav, Nuno Clara, Janice Eberly, Christian Goulding, Anne Hansen, Songrun He, Kilian Huber, Wei Jiang, Sophia Kazinnik, Tim Loughran, Linying Lv, Yueran Ma, Chuck Manski, Suproteem Sarkar, Andrei Shleifer, Junbo Wang, Daniel Weitz, and participants in Duke's AI Reading Group and JOIM AI in Finance Conference, INFORMS Conference on Financial Engineering and Fintech, as well as seminar participants at the Virtual Corporate Finance Conference, the College of William & Mary, and the FEB-RN and FMA International Global Seminar for helpful comments. Correspondence: Campbell R. Harvey, Duke University, Fuqua School of Business, 100 Fuqua Drive, Durham, NC 27708. Email: cam.harvey@duke.edu.

Chief Financial Officers occupy a unique vantage point in the economy. Operating at the intersection of corporate strategy and the macroeconomic environment, CFOs aggregate information about customers, suppliers, competitors, and capital markets into forward-looking expectations of business activity. Research spanning a quarter century has shown that measured CFO expectations, often gathered via surveys, are economically meaningful in predicting corporate outcomes.[1] However, despite their informational value, managerial expectations are expensive to collect (Bloom and Van Reenen, 2007), available only at a fixed frequency, and are limited to the companies whose executives choose to participate in a given survey. In this paper, we explore whether it is possible to rapidly generate credible CFO-style expectations at scale, at low cost, and without the delay of a human survey.

We propose and evaluate a method for generating synthetic CFO expectations with large language models (LLMs). We investigate how well an LLM, given an information set that includes a company profile and a CFO respondent's own prior answers to a survey of expectations, can predict what a respondent will report at a given point in time. It is important to be clear that our paper is *not* about exploring how accurately LLMs predict realizations of macroeconomic variables or corporate outcomes; rather, the object we forecast is the CFO's stated expectation at a given point in time. Even if CFO expectations are biased and imperfect predictors of the future (Ben-David et al., 2013; Malmendier and Tate, 2015; Barrero, 2022; Boutros et al., 2026), they are meaningfully associated with important corporate decisions and therefore are valuable in understanding corporate choices and outcomes. We also note that we make no claim that the model reasons in the same manner as the matched executive.

We prompt the LLM to role-play as the CFO of a specific public company at a specific point in time, supplying the model with the firm's industry, revenue, headcount, and geographic footprint, and imposing an information cutoff matching the date the actual survey was administered. We ask the model to answer a question from the Duke-Federal Reserve CFO Survey: "*Rate your optimism about the overall U.S. economy on a scale from 0–100, with 0 being the least optimistic and 100 being the most optimistic.*" When the same company's CFO has responded in earlier survey waves, we also supply that person's own prior optimism scores and the model's own prior forecasts for the same person, always restricted to dates before the survey cutoff.

[1] These outcomes include investment (Gennaioli et al., 2016), inflation (Fried et al., 2026), revenue and wages (Baslandze et al., 2026), planning (Graham, 2022) and macroeconomic variables (Coibion et al., 2018). Quarterly survey reports through 2020Q1 are archived at `https://cfosurvey.fuqua.duke.edu` and reports since then at `https://www.cfosurvey.org`.

We generate LLM forecasts for each of 6,075 survey responses from public companies over 2002–2025 and match every synthetic score to the response of the actual CFO at the same firm in the same quarter. Because our panel links responses to specific individuals over time, we can perform our analysis at the *individual* level. In contrast, previous work compares the *distribution* of synthetic answers with the distribution of human responses (e.g., Hansen et al., 2025). In our analysis, we find that the LLM reasonably predicts individual CFO responses produced by its human twin, as well as at the aggregate level.

A key contribution relative to earlier persona-based work is that we condition on each CFO respondent's own history. Models tuned on human feedback are known to compress the dispersion of human opinion: Santurkar et al. (2023) find that such models match the modal view of a group rather than the spread of views within it. Supplying the model with a person's prior answers anchors the persona to the individual rather than to a group or a generic firm type, and we find that this limits the collapse of the distribution. In our sample, the within-quarter dispersion of LLM scores is about 75% of the dispersion in the human panel.

Our experimental design deals with three potential concerns of persona-based LLM studies. First, the LLM might access information after the date that the human CFO completed the survey. We address this by imposing information cutoffs that instruct the model not to access post-survey data, and any respondent history we supply is itself dated strictly before the cutoff. We find supportive evidence that this instruction was followed in that the first-quarter 2020 LLM forecast is not contaminated by the COVID-era reduction in optimism that occurred ex post.

A second potential concern relates to whether the target variable itself (the individual CFO optimism score produced at a given point in time) is known ex ante to the model. In our experiment, the variable of interest was never part of the model's training corpus because (i) CFO-specific expectations are never released publicly, and (ii) in our application, the survey data are submitted through Duke University's institutional agreement with OpenAI and are contractually excluded from training. The model cannot recall what it has never seen. Our setting thus satisfies the no-leakage condition formalized by Ludwig et al. (2026) (see also Gao et al., 2026).

A third potential concern is that the LLM is not forecasting per se but instead is just repeating the human CFO's expectation from a previous quarter. We address this by explicitly controlling for the lagged human response and find that even controlling for the lagged human expectation, the LLM forecast has a significant relation with the human CFO's current-period expectation.

We explore whether our main findings hold for alternative firm-specific expectation variables. Repeating the exercise for the respondent's optimism about the financial prospects of their own firm yields a fit similar to the main analysis. We also ask the LLM to predict the firm-quarter expectation of firm-specific revenue growth; the relation remains positive but is statistically weaker. We also show that our main results are robust to an alternative open-weight model.

Next, we explore how the predictive accuracy of the LLM forecast varies conditional on the information set. We find that the fit increases monotonically with the amount of information provided about a given respondent's own history of past expectations. A parallel 'more information leads to better forecasts' pattern holds for the breadth of firm characteristics provided to the LLM. The most important information is the respondent's own history of past expectations.

Horton et al. (2023) introduce the concept of *Homo silicus*, the idea that LLMs can be treated as computational models of humans, capable of being assigned endowments, information, and roles and then interviewed at scale (see also Manning et al., 2024; Aher et al., 2023).[2] The work closest to ours is Hansen et al. (2025), who construct synthetic Survey of Professional Forecasters (SPF) personas and show that LLM-generated forecasts replicate key distributional properties of the human survey. We go further by testing whether the LLM persona predicts the response of the *specific* human at the *same* firm in the same quarter, and by showing how that fidelity depends on the information the model is given.[3]

The remainder of the paper proceeds as follows. Section I describes the construction of the LLM optimism scores, the respondent-history conditioning, and the data sample. Section II presents the empirical analysis. Section III summarizes the findings, discusses limitations, and outlines possible applications of CFO digital twins.

[2]See also Argyle et al. (2023), Suh et al. (2025), Mei et al. (2024), and Manning and Horton (2026).

[3]In a non-economic setting, Park et al. (2026) build generative agents from qualitative interviews and show that the agents reproduce those same individuals' survey responses with high fidelity. Relative to their design, we do not require access to the respondent beyond observable firm characteristics and the respondent's own panel of past answers, and we validate against high-stakes professional expectations rather than general social attitudes. Chetty et al. (2026) use "digital twins" in a related sense, building machine-learning counterfactuals from administrative data and validating them against randomized experiments.

## I. Construction of the LLM Optimism Scores

We construct LLM-generated forecasts to simulate the expectations of individual CFOs participating in The CFO Survey, conducted quarterly by Duke University and the Federal Reserve Banks of Atlanta and Richmond (Graham et al., 2020). Our approach employs a large language model to role-play as financial executives at specific companies at a specific date each quarter, incorporating contextual information available at each forecast date. We start with a survey response by a human CFO, to which we append a detailed firm-executive profile and a history of responses across survey waves from this same CFO; see Internet Appendix A for additional details.

### *A. LLM Response Generation*

We access the model through Duke's AI Gateway, which routes requests to an Azure-hosted deployment of the model under Duke's data-security agreement, using OpenAI's Responses API and the model `GPT-5.4`. All results in the paper come from a single set of runs executed beginning April 21, 2026 (see Internet Appendix D on reproducibility). Accessing OpenAI via Duke's institutional agreement ensures that submitted survey data are not used for model training; accordingly, the model does not retain or learn the contents of these surveys across runs.

For each actual human survey response in our sample, we create a corresponding LLM-generated response using two prompts: a *system prompt* that establishes the context, persona, and analytical instructions, and a *user prompt* that provides the firm-specific profile, the respondent's history (when available), and that poses the survey question.

*(i) System Instructions.* We provide the model with detailed instructions that describe the context of The CFO Survey and the forecasting task (see Internet Appendix B). In particular, the model is instructed to use only the information available as of the date the survey was sent out (i.e., excluding all information after the cutoff). This temporal constraint attempts to ensure that the LLM operates under the same information set as the human CFO at the time of their response, and to address the look-ahead concerns raised by Lopez-Lira et al. (2025) and Fritsch and Mazlish (2026), who show that without such constraints, LLM forecasts may partially recall outcomes from training data (see also Wu et al., 2025). We also enable the model's web-search tool, subject to the same cutoff, so that it can draw on the kinds of public information a CFO would have had at the time, such as earnings reports, CFO interviews, and analyst commentary.

Although it is only a single quarter, the 2020Q1 survey provides an informal test for look-ahead bias. The survey was conducted just weeks before the COVID-19 pandemic triggered a global economic collapse, and in this test we restrict the model's information cutoff to February 14, 2020. Even with web search enabled, the LLM's expectations for this quarter exhibit no reduction in optimism relative to immediate prior quarters, mirroring the sentiment of human forecasters who similarly had little forewarning of the impending crisis. In this particular quarter, our prompting strategy successfully prevents the model from incorporating knowledge of subsequent events: a model with access to post-cutoff information would almost certainly produce far more pessimistic forecasts for this period.

The system prompt also lists the information we ask the model to consider when forming its forecast. Specifically, we instruct the model to consider historical financial performance (3–5 years of revenue data, growth trends, profit margins), forward-looking company indicators (management guidance, product pipelines, order backlogs, expansion plans), analyst coverage (consensus estimates, price targets, key assumptions), comparable-company analysis where direct company information is limited, market and industry context (growth projections, competitive dynamics, regulatory changes), and macroeconomic factors (GDP forecasts, interest rates, inflation, and sector-specific indicators).

*(ii) Company-Specific Profile.* For each response, we provide a profile containing all available information about the responding executive and their firm as of the survey cutoff date (see Internet Appendix C for a sample query). This profile includes the executive's name, job title, company name, ticker, industry classification, prior-year sales revenue, total employment, percentage of foreign revenue, headquarters location, state, ownership status, and credit rating. When certain identifying fields (job title, company name, industry, and headquarters location) are missing for a given wave, we impute them using the respondent's nearest survey response in time, as determined by the survey completion date. That said, the LLM has access to this information from the web search. We use survey information to fill in this information because in one experiment we turn off the Web search. All remaining missing fields are simply omitted from the prompt (see Internet Appendix A).

*(iii) Respondent History.* Whenever a respondent has answered the survey in earlier waves, we supply the model with that person's own history, strictly limited to dates before the survey cutoff. For each respondent, we inject (a) up to 12 recent prior optimism scores the person reported (within the preceding 12 calendar quarters), (b) the model's own prior forecasts for the same person over the same window, and (c) the person's average optimism, computed only from quarters provided in (a). When a respondent has no usable history, this section of the prompt is left blank. The system prompt asks the model to treat this history as a strong prior on the individual's style, optimism bias, and level (i.e., how they use the 0–100 scale and where they typically answer), unless information available at the cutoff date clearly points in a different direction (Internet Appendix B). Table 2, Panel C, reports the exercise with the history block removed.

*(iv) Survey Question.* As mentioned earlier, the specific survey question we ask is: "Rate your optimism about the overall U.S. economy on a scale from 0–100, with 0 being the least optimistic and 100 being the most optimistic."

Throughout the paper, "optimism" means optimism about the overall U.S. economy. In a robustness check, when we refer to the respondent's assessment of their own company, we say "own-firm optimism". In Section II.C, we test own-firm optimism as well as expected revenue growth, to determine whether our findings hold more broadly.

### *B. Implementation and Stability*

For each of the 6,075 survey responses in our sample, we ask the LLM to produce a forecast three separate times using the same input, which gives 18,225 forecasts (6,075 × 3) in total. We refer to each of these as a "call" and to the set of three as "replications". LLM outputs are not deterministic, so identical prompts can yield different answers across calls; the three replications let us measure how far the forecast changes from call to call.

Table IA2 in Internet Appendix D reports stability statistics for the three replications. We find that roughly 96% of total variance is attributable to differences across survey observations rather than stochastic variation within repeated calls, indicating that the three forecasts are very highly correlated. Additional details provided in the appendix reinforce the broader finding that LLM-generated survey data can achieve high internal consistency (Argyle et al., 2023; Li et al., 2024). For the remainder of the paper, we use the mean of the three calls as the LLM forecast for each survey response.

### *C. Macro Forecasts and Consumer Sentiment*

We also examine the performance of the LLM model after aggregating all responses in a given quarter because policy-makers often care about aggregate projections of business sector ac-tivity. We compare these economy-wide LLM aggregates against two established measures of macroeconomic expectations. The first is the mean year-ahead real GDP growth forecast (`drgdp5`) from the Survey of Professional Forecasters (SPF), administered quarterly by the Federal Reserve Bank of Philadelphia and drawing on roughly 30 to 40 professional forecasters. The second is the University of Michigan Index of Consumer Sentiment (`ics_all`), based on interviews with about 500 households, calibrated to a value of 100 in 1966. Consumer sentiment carries forward-looking information about consumption and aggregate demand (Ludvigson, 2004), which makes it a natural complement to a survey of companies. Section II.E uses both to determine whether the LLM tracks CFO optimism beyond what these gauges already capture.

### *D. Summary Statistics*

Table 1 presents summary statistics for the main variables. Panel A reports individual-level statistics over the full sample of 6,075 matched responses. Mean LLM optimism (56.6) is lower than mean CFO Survey optimism (60.1) produced by humans, though the two distributions have comparable dispersion at the individual level (standard deviations of 16.2 and 17.6, respectively).

Panel B reports cross-sectional disagreement at the quarterly level. In each quarter, we compute the standard deviation of optimism scores across all respondents; the table reports the time-series distribution of these within-quarter dispersions. The average within-quarter standard deviation of CFO Survey responses is 16.5, compared to 12.3 for the LLM.[4] The disagreement ratio, the within-quarter LLM standard deviation divided by the within-quarter CFO standard deviation, averages 0.75 (median of 0.74), indicating that the synthetic panel under-disperses by roughly 25%. Instruction-tuned models are known to collapse toward modal opinions (Santurkar et al., 2023), a problem that persists even after explicitly steering models toward specific demographic profiles (also see Vafa et al., 2024).[5] In our sample, this compression is inversely related to the amount of historical data in the information set: the average disagreement ratio is 0.26 for first-time respondents versus 0.90 for respondents whose prior answers appear in the prompt.

[4] Our use of within-quarter cross-sectional dispersion as a disagreement measure connects to a broader literature that analyzes the relation between disagreement proxies; see Goulding et al. (2026a,b).

[5] Hansen et al. (2025) present related evidence. They simulate the Survey of Professional Forecasters with personas built from hand-collected forecaster characteristics, but because SPF responses are anonymized they cannot condition each persona on that individual's own past answers. The individual-level forecast distributions are visibly more concentrated than the human distributions in most of the quarters and variables they report.

**Table 1:** Summary Statistics

The sample consists of 6,075 matched observations from the Duke–Federal Reserve CFO Survey (Graham and Harvey, 2001; Graham et al., 2020) of public-company CFOs collected between 2002 and 2025. *Panel A* reports individual-level statistics over the full sample. *Panel B* reports cross-sectional disagreement: for each of the 93 quarters in the sample, we compute the standard deviation of optimism scores across all respondents and report the time-series distribution of these within-quarter dispersions; the disagreement ratio is the within-quarter LLM standard deviation divided by the within-quarter CFO standard deviation, computed quarter by quarter. *Panel C* reports time-series variation by aggregating observations for each quarter enabling comparison with macroeconomic benchmarks available at that frequency. LLM Optimism and CFO Survey Optimism are measured on a 0–100 scale; Michigan Consumer Sentiment is an index level; SPF Real GDP Growth is the Survey of Professional Forecasters consensus forecast for one-year-ahead real GDP growth, in percent.

| | Obs | Mean | Median | SD |
|---|---|---|---|---|
| *Panel A: Full Sample (Individual Level)* | | | | |
| LLM Optimism | 6,075 | 56.57 | 59.33 | 16.19 |
| CFO Survey Optimism | 6,075 | 60.13 | 60.00 | 17.62 |
| *Panel B: Cross-Sectional Disagreement (Quarterly)* | | | | |
| LLM Optimism (within-quarter SD) | 93 | 12.31 | 12.21 | 3.04 |
| CFO Optimism (within-quarter SD) | 93 | 16.53 | 16.09 | 2.82 |
| Disagreement Ratio (LLM/CFO) | 93 | 0.75 | 0.74 | 0.16 |
| First-time respondents | 70 | 0.26 | 0.22 | 0.16 |
| Respondents with history | 92 | 0.90 | 0.90 | 0.20 |
| *Panel C: Quarterly Aggregates and Time-Series Variation* | | | | |
| LLM Optimism | 93 | 58.61 | 60.33 | 9.47 |
| CFO Survey Optimism | 93 | 60.78 | 61.09 | 6.70 |
| Michigan Consumer Sentiment | 93 | 80.53 | 81.60 | 12.78 |
| SPF Real GDP Growth | 93 | 2.70 | 2.72 | 0.75 |

Panel C reports time-series variation after first aggregating observations to the quarterly level (93 observations). Quarterly mean LLM optimism (58.6) and CFO Survey optimism (60.8) are similar in level, although the LLM aggregate is somewhat more volatile across quarters (standard deviation of 9.5 versus 6.7). The Michigan Consumer Sentiment Index is scaled differently and the SPF provides a growth forecast rather than a level.

## II. Main Analysis and Results

### *A. Do LLMs Track Individual CFO Expectations?*

We first explore whether the forecast of the digital twin is close to the expectation of the human CFO at the same firm in the same quarter. This is a person-specific, matched test[6] that goes well beyond the distributional validation currently in the literature: rather than asking whether the synthetic population looks like the real distribution in the cross-section, we ask whether the digital twin, given the firm's observable characteristics, a temporal information cutoff, and the respondent's own pre-cutoff history, can approximate a specific human executive's expectation.

In Table 2, the dependent variable is the individual CFO's optimism score, and the key regressor is the matched LLM score. In every specification the LLM score is positively and highly significantly associated with the human optimism score. In the baseline (column 1), the LLM score significantly predicts the CFO's reported optimism ($t = 18.0$), and accounts for 27% of the total variation in individual responses.

Adding firm fixed effects (column 2) absorbs each firm's time-invariant average optimism.[7] Because the survey is typically answered by a single executive per firm, the firm fixed effect effectively differences out that respondent's own long-run mean, so the coefficient is identified from *within-firm, over-time* variation (equivalently, movements in a given CFO's optimism around their personal average). Estimating off this variation asks whether the model can track when a particular CFO becomes more or less optimistic than usual. In this analysis, the LLM score remains large and significant ($t = 12.9$), so its predictive content is not only the result of persistent differences across firms or executives.

Adding year-quarter fixed effects instead (column 3), which absorb all variation common across CFOs within a given quarter, also leads to a highly significant relation ($t = 16.7$), implying that the model discriminates across CFOs *within* the same quarter and not merely across quarters. The most demanding specification (column 4) includes firm and year-quarter fixed effects simultaneously, isolating, for each CFO, deviations from their own average that are not also shared by other executives that quarter; even here, the LLM output forecasts the human CFO's optimism in a significant manner ($t = 6.7$). This specification explains nearly half the variation in individual optimism.

[6] Of the 1,218 firms in our sample, 38 (3%) have responses from two different named individuals over the sample period. For the remaining 97%, a firm-quarter match is also a match to the same person. Respondent history is constructed at the person level (Internet Appendix A), so the history supplied to the model belongs to the individual answering in a given quarter even in these 38 firms.

[7] The sample size falls from 6,075 to 5,631 observations in the firm fixed-effect specifications because 444 observations belong to firms that appear in only one survey wave. The same reason explains the 1,980 to 1,818 reduction in sample size in columns (6) and (8), and for the 3,832 to 3,624 drop in Table IA3.

Columns (5)–(8) address an obvious concern about our history-conditioning design: the LLM might simply echo the respondent's own recent response, which is one of the inputs we supply. In this analysis, we restrict the sample to observations for which the respondent answered in the immediately preceding survey wave; and, we include that prior answer (*Last-Quarter CFO Optimism*) as a control. The LLM score remains a strong and significant predictor in all four specifications, with $R^2$ above 0.50. The model therefore interprets historical information beyond the linear persistence in the respondent's previous answer. Once fixed effects are added, the prior answer is largely subsumed, while the LLM score is not. That the LLM prediction tracks the human response even after firm and year-quarter fixed effects and a control for the prior answer suggests that the model captures individual CFO expectations.

### *B. More Information, Better Predictions*

If the LLM is using the information we supply, accuracy should improve as we give the model more information to work with. Panel B of Table 2 splits the sample two ways to test this. Columns (1)–(3) of Panel B split the sample by how much of the respondent's own history the model receives. When no prior responses are available, typically a respondent's first appearance in the survey, the LLM score predicts optimism but explains only about 10% of the variation. When one to three prior data points are available to the model, the fit more than triples ($R^2 = 0.33$), and when more than three prior responses are supplied to the model, it increases to nearly half ($R^2 = 0.49$). The model appears to use the respondent's track record to sharpen its individual-level forecasts of the respondent's optimism score.

Columns (4)–(5) of Panel B split the sample by the breadth of firm characteristics included in the prompt. Not every survey wave records all 12 profile fields, and the median prompt contains 10 characteristics, which is the threshold we use to divide the sample. The full list of profile fields is provided in Internet Appendix C. When fewer than 10 are available, the

**Table 2:** Predicting Individual CFO Survey Optimism About the Overall Economy

This table reports OLS coefficients from regressions of individual CFO Survey optimism on the matched LLM optimism score. *Panel A* adds firm and year-quarter fixed effects, first on all matched observations and then on respondents who also answered the optimism question in the preceding survey wave, controlling for that prior response (*Last-Quarter CFO Optimism*). *Panel B* reports baseline regressions within mutually exclusive, exhaustive subsamples defined by the respondent's number of prior responses and by the number of firm characteristics supplied in the prompt. *Panel C* re-generates the LLM optimism score under six prompt designs. Each design includes or removes three blocks of conditioning information: the *firm profile* (all profile fields in Internet Appendix C, included or removed as one block), *web retrieval*, and the *respondent's own history*. Column (6) is the full prompt. Panels B and C include no fixed effects. Standard errors are two-way clustered at the firm and year-quarter levels (Cameron et al., 2011); $t$-statistics in parentheses. *, **, and *** denote significance at the 10%, 5%, and 1% levels assuming a single hypothesis test.

*Panel A: Baseline and Prior-Response Control; Dep. variable = CFO Survey Optimism*

| | No History Control | | | | Control for Last-Quarter Resp. | | | |
|---|---|---|---|---|---|---|---|---|
| | (1) | (2) | (3) | (4) | (5) | (6) | (7) | (8) |
| LLM Optimism | 0.565*** | 0.429*** | 0.578*** | 0.276*** | 0.521*** | 0.527*** | 0.489*** | 0.286** |
| | (17.98) | (12.86) | (16.72) | (6.67) | (7.50) | (7.04) | (3.94) | (2.33) |
| Last-Quarter CFO Optimism | | | | | 0.217*** | 0.0366 | 0.251** | 0.171* |
| | | | | | (3.04) | (0.51) | (2.21) | (1.84) |
| Firm FE | | Y | | Y | | Y | | Y |
| Year-Quarter FE | | | Y | Y | | | Y | Y |
| $N$ | 6,075 | 5,631 | 6,075 | 5,631 | 1,980 | 1,818 | 1,980 | 1,818 |
| $R^2$ | 0.269 | 0.446 | 0.305 | 0.494 | 0.502 | 0.612 | 0.547 | 0.662 |

*Panel B: Predictive Accuracy by Amount of Information*

| | Respondent History | | | Firm Characteristics | |
|---|---|---|---|---|---|
| | None (1) | 1–3 (2) | >3 (3) | <10 (4) | 10–12 (5) |
| LLM Optimism | 0.367*** | 0.604*** | 0.743*** | 0.535*** | 0.585*** |
| | (8.65) | (23.29) | (19.62) | (17.08) | (14.56) |
| $N$ | 2,237 | 2,236 | 1,602 | 3,102 | 2,973 |
| $R^2$ | 0.100 | 0.334 | 0.494 | 0.214 | 0.317 |

*Panel C: Predictive Accuracy by Prompt Design*

| | (1) | (2) | (3) | (4) | (5) | (6) |
|---|---|---|---|---|---|---|
| LLM Optimism | 0.350*** | 0.368*** | 0.354*** | 0.568*** | 0.604*** | 0.646*** |
| | (7.28) | (6.96) | (7.28) | (21.85) | (22.39) | (27.86) |
| Firm profile | | Y | Y | | Y | Y |
| Web retrieval | Y | | Y | | | Y |
| Respondent history | | | | Y | Y | Y |
| $N$ | 3,594 | 3,594 | 3,594 | 3,594 | 3,594 | 3,594 |
| $R^2$ | 0.095 | 0.095 | 0.096 | 0.308 | 0.357 | 0.375 |

LLM score explains about 21% of the variation, whereas prompts containing ten to twelve characteristics achieve a better fit ($R^2 = 0.32$). The results support the idea that supplying additional information improves the LLM's ability to forecast the CFO responses.

The subsample splits in Panel B hold the prompt fixed and vary the sample. Panel C of Table 2 does the reverse, holding the sample size fixed and re-generating every forecast with one or more blocks of conditioning information removed from the prompt. The three blocks are the firm profile, web retrieval, and the respondent's own history. In Panel C the firm profile means the full set of profile fields in Internet Appendix C, which we either include or remove as a whole. This is different from the rightmost columns in Panel B, where every prompt contains the profile and we split the sample by how many of its 12 fields are non-missing for a given respondent.

The respondent's own history is by a wide margin the most valuable input (Panel C). Prompts built from the firm profile alone, from web retrieval alone, or from the two together explain about 10% of the variation in individual responses. A prompt containing only the respondent's own history explains 31%. Adding the firm profile to that history raises the fit to 36%, and the full prompt, which also enables web retrieval, reaches 38%. Varying the prompt therefore leads to the same conclusion as Panel B: the more information the model receives, the more closely its forecast tracks the human response.

### *C. Other Survey Questions: Own-Firm Optimism and Revenue Growth*

The U.S.-economy optimism question is the item with the longest consistent history in The CFO Survey, which is one reason we focus on it above. Our empirical approach, however, is not specific to this question. Specifically, we repeat the exercise for two additional questions that have similarly long histories: the CFO's optimism about the financial prospects of *their own company* (also on a 0–100 scale), and the CFO's expected revenue growth for their own firm over the next 12 months. Own-firm optimism is a narrower and more inward-looking judgment than optimism about the aggregate economy, and expected revenue growth is a rate of change of a common planning variable rather than a sentiment score level; together they test the empirical design along two different dimensions. The prompts, the information cutoffs, and the respondent-history conditioning are constructed analogous to that in Section I.

Table 3 reports the results. Panel A uses own-firm optimism as the dependent variable. In the baseline (column 1) the LLM own-firm optimism score is highly significant ($t = 25.7$), accounting for 24% of the variation, similar to what we obtained for U.S.-economy optimism in Table 2. The coefficient remains highly significant once firm and year-quarter fixed effects are imposed together ($t = 8.6$). Columns (3)–(6) add the respondent's own prior answer to the same question as a control, first for the subsample with any prior response on record and then for the narrower subsample that answered in the immediately preceding wave. The pattern here is sharper than for U.S.-economy optimism: once the LLM score is included, the respondent's prior own-firm optimism is statistically indistinguishable from zero in every column. Estimated on its own, the prior answer is a strong predictor in both subsamples (not reported in a table), but when both are included together, the LLM score subsumes it. The digital twin's forecast already captures whatever (potentially non-linear) persistence there is in an executive's lagged optimism about their own firm.

Panel B examines how well the digital twin forecasts firm-quarter matched expected revenue growth. The fit is weaker than for either optimism measure, which is unsurprising: a point forecast of a firm's own revenue growth is much more volatile than the optimism level. Still, the LLM forecast has a positive and significant relation with the CFO's revenue expectation in the baseline and in the prior-response specification, and it survives the saturated fixed-effect specification in columns (2) and (4). In the smallest subsample, those respondents who answered in consecutive waves, the coefficient loses significance.

*D. Heterogeneity in Fit Across Quarters*

In this section, we estimate the $R^2$ from a cross-sectional regression of CFO optimism on the matched LLM score *separately within each of the 93 quarters*. The distribution of $R^2$ values is wide, ranging from close to zero in a few early quarters to above 0.90 in the best-fitting quarter. Consistent with the information-quantity results in Table 2, Panel B, the fit tends to improve in later quarters, when more respondents have accumulated the survey history that the model conditions on.

Figure 1 plots individual CFO optimism against the matched LLM score for four representative quarters: the 25th, 50th, 75th, and 100th percentiles of the quarterly $R^2$ distribution. The relation is positive in every panel, including the 25th percentile. The panels also distinguish respondents by information set. Cyan squares identify CFOs for whom the model conditions on prior survey history, and black circles identify first-time respondents. The "first-timer" points in the 50th and 75th percentiles cluster in a narrow vertical band, indicating that without

**Table 3:** Predicting Other CFO Survey Responses: Own-Firm Optimism and Revenue Growth

This table repeats the design of Table 2 for two additional items on the Duke–Federal Reserve CFO Survey. The dependent variable is the respondent's optimism about the financial prospects of their own company, on a 0–100 scale (*Panel A*), and their expected revenue growth for their own firm over the next twelve months, in percent (*Panel B*). *LLM Own-Firm Optimism* and *LLM Revenue Growth* are the matched LLM forecasts for the respective items, generated with the prompts, information cutoffs, and respondent-history conditioning of Section I. Columns (1)–(2) use all matched observations; columns (3)–(4) restrict to respondents with at least one prior response to the same item on record; columns (5)–(6) restrict further to those who answered in the immediately preceding wave. *Prior Response* is the respondent's own previous answer to the same item. The even-numbered columns include firm and year-quarter fixed effects simultaneously. Standard errors are two-way clustered at the firm and year-quarter levels; $t$-statistics are in parentheses. *, **, and *** denote significance at the 10%, 5%, and 1% levels, respectively, assuming a single hypothesis test.

*Panel A: Dep. variable = Own-Firm Optimism (0–100)*

| | No History Control | | Any Prior Response | | Last-Quarter Resp. | |
|---|---|---|---|---|---|---|
| | (1) | (2) | (3) | (4) | (5) | (6) |
| LLM Own-Firm Optimism | 0.604*** | 0.321*** | 0.644*** | 0.409*** | 0.694*** | 0.408*** |
| | (25.65) | (8.64) | (11.58) | (6.21) | (7.77) | (3.35) |
| Prior Response | | | −0.011 | −0.038 | 0.017 | −0.009 |
| | | | (−0.26) | (−0.85) | (0.20) | (−0.10) |
| Firm FE | | Y | | Y | | Y |
| Year-Quarter FE | | Y | | Y | | Y |
| $N$ | 6,038 | 5,588 | 3,803 | 3,595 | 1,988 | 1,831 |
| $R^2$ | 0.240 | 0.424 | 0.315 | 0.475 | 0.409 | 0.580 |

*Panel B: Dep. variable = Expected Revenue Growth (%)*

| | No History Control | | Any Prior Response | | Last-Quarter Resp. | |
|---|---|---|---|---|---|---|
| | (1) | (2) | (3) | (4) | (5) | (6) |
| LLM Revenue Growth | 0.451*** | 0.219* | 0.524*** | 0.396*** | 0.255 | 0.193 |
| | (3.72) | (1.71) | (2.95) | (2.71) | (1.09) | (0.79) |
| Prior Response | | | 0.083 | 0.000 | 0.059 | 0.033 |
| | | | (1.16) | (0.00) | (1.66) | (0.69) |
| Firm FE | | Y | | Y | | Y |
| Year-Quarter FE | | Y | | Y | | Y |
| $N$ | 3,045 | 2,646 | 1,611 | 1,493 | 901 | 823 |
| $R^2$ | 0.071 | 0.361 | 0.104 | 0.165 | 0.148 | 0.311 |

respondent-specific history the model generates little cross-sectional spread. This is consistent with our results in Table 1 that first-time respondents offer very little dispersion of forecasts. There is an upward-sloping fit for the subsample of CFOs with past information (the square cyan points).

### *E. Aggregating to the Quarterly Level*

We also explore the relation between LLM and CFO optimisms once the panel structure is collapsed down to the quarterly frequency by averaging the data within each year-quarter. This allows us to test whether the LLM tracks movements in average CFO optimism over time in the absence of cross-sectional variation within a quarter. We benchmark against the Michigan Consumer Sentiment Index and the SPF one-year-ahead real GDP growth forecast. Table 4 reports regressions of quarterly mean CFO optimism on quarterly mean LLM optimism and the two benchmarked series.

The mean LLM score enters positively and significantly when paired with the Michigan Consumer Sentiment index (column 1) and when paired with the SPF forecast (column 2). On their own, the two human benchmarks explain much less of the variation in quarterly CFO sentiment (column 3). With all three measures in the same specification (column 4), the LLM score remains highly significant. Neither Michigan sentiment nor the SPF forecast is significant once the LLM score is included. The LLM score appears to capture variation in aggregate executive optimism that goes beyond these standard measures. This aggregate result also indicates that it is unlikely that the individual-level findings are an artifact of the panel structure. Column (5) repeats the full specification for own-firm optimism, where the quarterly mean LLM forecast enters with a highly significant coefficient; the SPF forecast is also moderately significant in this column, while Michigan sentiment is not. Column (6) does the same for expected revenue growth over the 71 quarters in which that item is available. Aggregate LLM revenue growth is significantly related to aggregate CFO-provided expected revenue growth.

The results in Table 4 are robust to using medians rather than means. In addition, Internet Appendix F reports a second robustness check along a different dimension, repeating the full exercise with a locally hosted open-weight model in place of GPT-5.4 and finding similar results.

**Figure 1:** LLM vs. CFO Optimism Across the Distribution of Quarterly Fit

Each panel is a scatter plot of individual CFO Survey optimism (vertical axis) against the matched mean LLM optimism score (horizontal axis) for a single survey quarter, with the OLS $R^2$ shown. *Cyan squares* denote respondents for whom the model conditions on prior survey history (i.e., the respondent has appeared in an earlier wave), while *black circles* denote respondents appearing in the sample for the first time, for whom no own-history is available. First-time respondents exhibit markedly lower dispersion in the LLM score within a given quarter, as the model defaults toward a narrower prior in the absence of respondent-specific information. The four quarters are chosen at the 25th, 50th, 75th, and 100th percentiles of the distribution of within-quarter $R^2$ values across the 93 sample quarters. Two $R^2$ are reported: overall (black) and for the subsample with history (cyan).

(A) 25th Percentile: 2020Q1

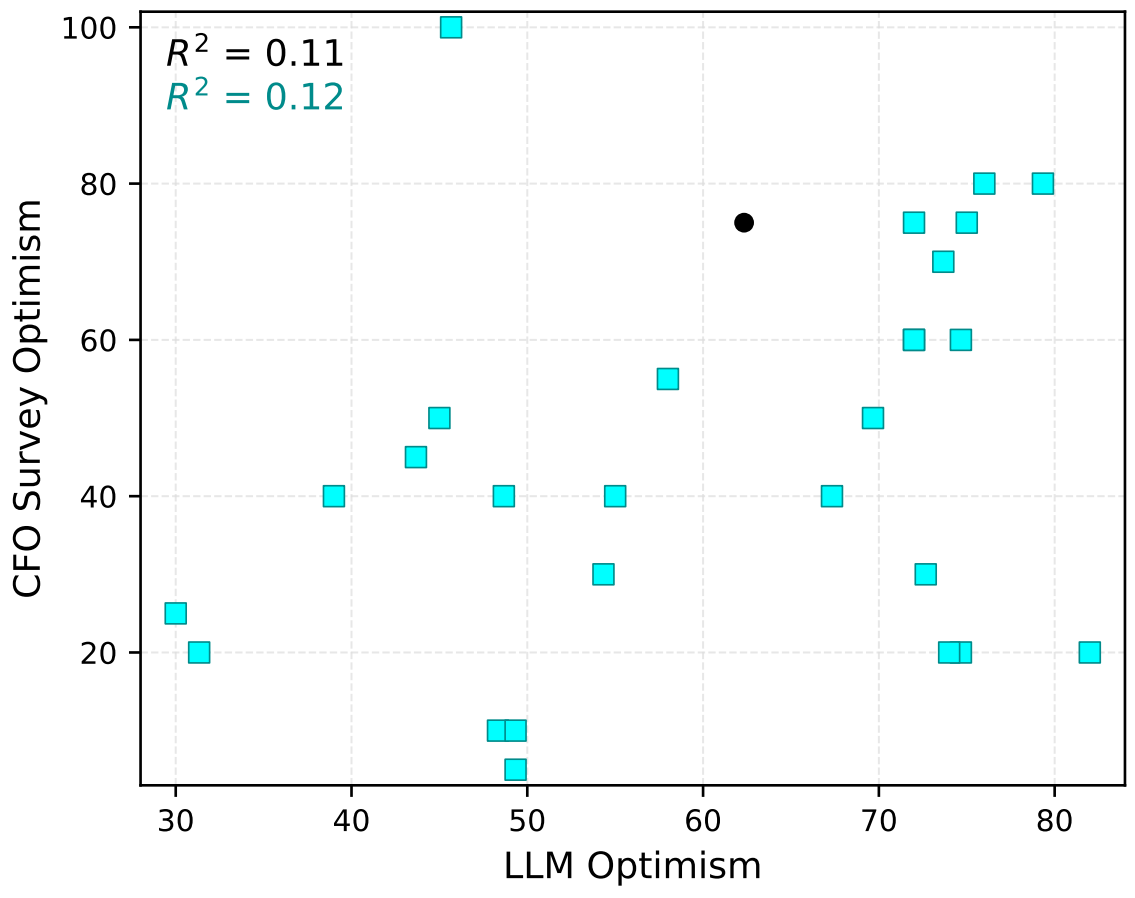

(B) 50th Percentile: 2015Q1

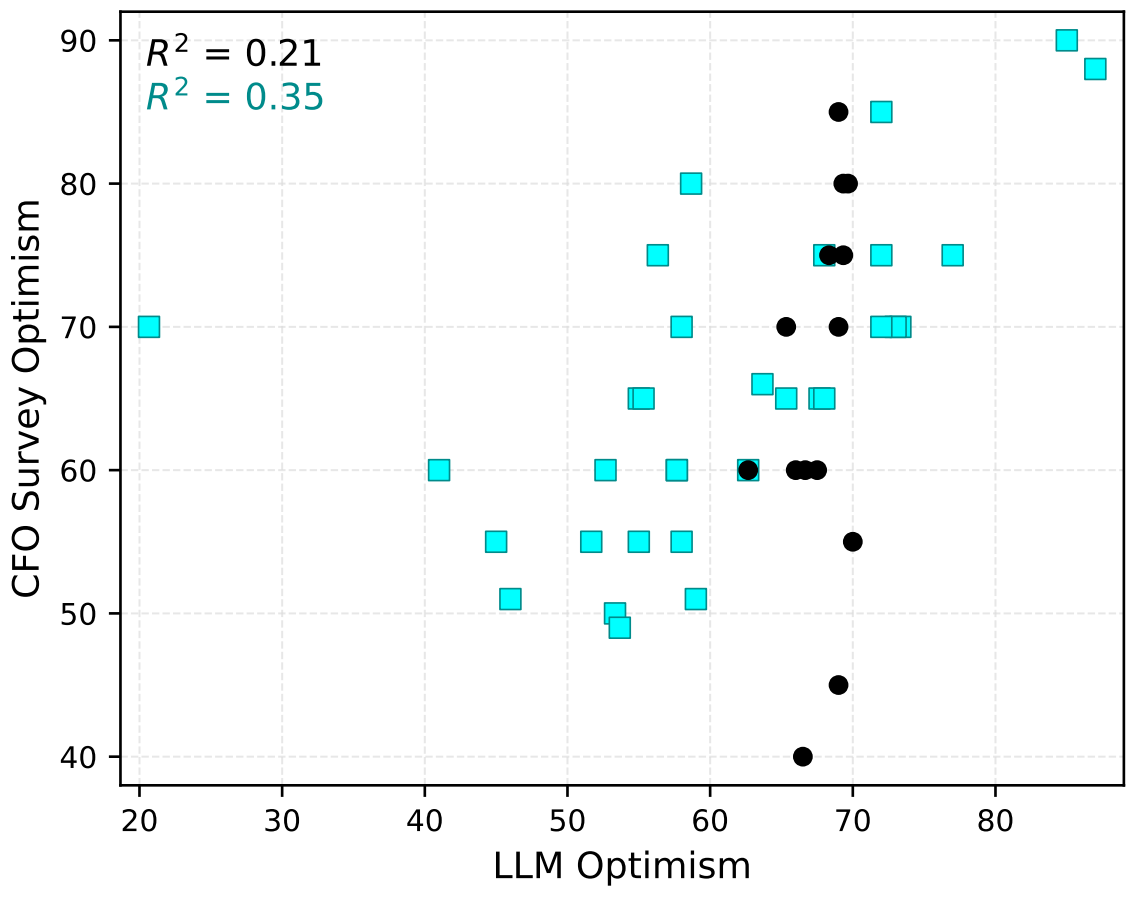

(C) 75th Percentile: 2014Q1

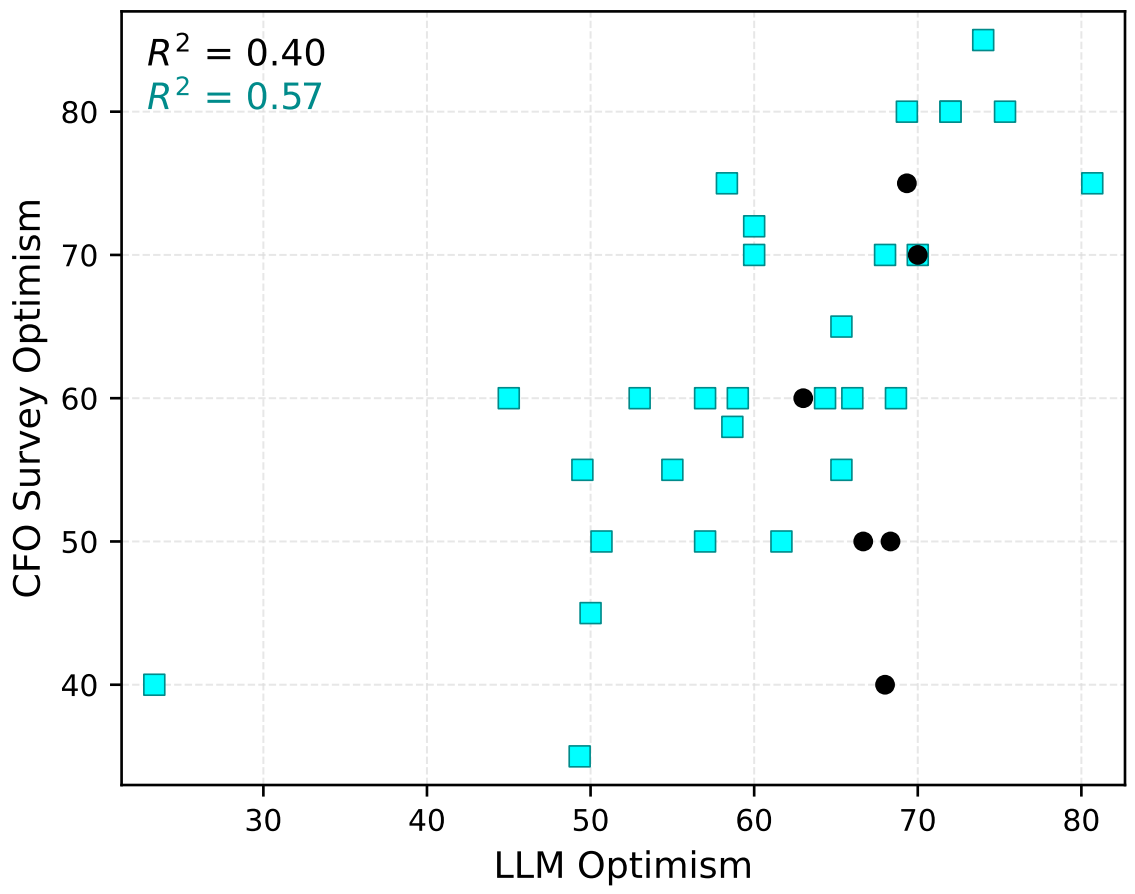

(D) 100th Percentile: 2022Q2

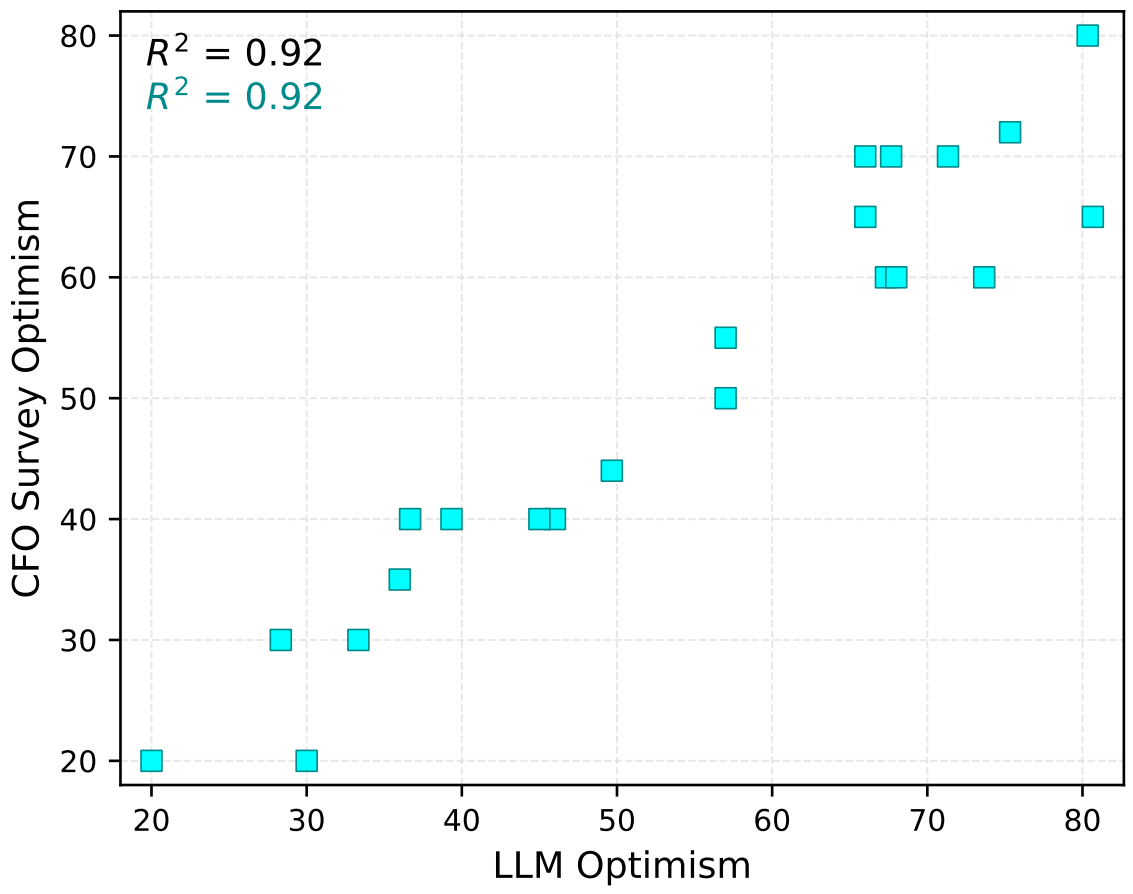

**Table 4:** LLM Forecasts and Quarterly CFO Survey Responses

This table reports quarterly-level OLS regressions of mean CFO Survey responses on the mean matched LLM forecast and macroeconomic benchmarks. The data are collapsed to one observation per year-quarter by averaging within quarter. In columns (1)–(4) the dependent variable is mean optimism about the overall U.S. economy. In column (5) it is mean own-firm optimism, and in column (6) it is mean expected revenue growth; in these two columns *LLM Forecast (Mean)* is the quarterly mean of the matched LLM forecast for the same item (Table 3). Column (6) covers the 71 quarters in which the revenue growth item is available. *Michigan Consumer Sentiment* is the index level; *SPF Real GDP Growth* is the Survey of Professional Forecasters consensus forecast for one-year-ahead real GDP growth. The sample spans 2002–2025. Standard errors are Newey–West (Newey and West, 1987) with 4 lags; $t$-statistics are in parentheses. *, **, and *** denote significance at the 10%, 5%, and 1% levels, respectively, assuming a single hypothesis test.

| | *CFO U.S. Economy Optimism* | | | | *Own-Firm Optimism* | *Revenue Growth* |
|---|---|---|---|---|---|---|
| | (1) | (2) | (3) | (4) | (5) | (6) |
| LLM Forecast (Mean) | 0.579*** | 0.586*** | | 0.574*** | 0.853*** | 0.521*** |
| | (7.11) | (8.82) | | (6.86) | (23.70) | (7.08) |
| Michigan Consumer Sentiment | 0.0228 | | 0.264*** | 0.0163 | −0.0354 | −0.0461 |
| | (0.47) | | (3.24) | (0.33) | (−0.96) | (−1.46) |
| SPF Real GDP Growth | | 0.479 | 1.161 | 0.432 | 1.516** | 0.419 |
| | | (1.18) | (1.10) | (0.98) | (2.59) | (0.84) |
| $N$ | 93 | 93 | 93 | 93 | 93 | 71 |
| $R^2$ | 0.715 | 0.716 | 0.316 | 0.717 | 0.777 | 0.326 |

## III. Summary and Discussion

Prompting an LLM to role-play as the CFO of a specific public company at a specific point in time, and supplying, if available, the respondent's own pre-cutoff history, we produce synthetic optimism scores matched to human CFO expectations spanning 2002–2025. Four main findings emerge.

First, the LLM score is a significant predictor of the actual CFO's response at the individual level. The predictive relationship survives firm fixed effects, year-quarter fixed effects, and a direct control for the respondent's most recent prior answer, indicating that the model contributes information beyond simple linear persistence.

Second, predictive accuracy improves with the information provided. Both the amount of a respondent's own history and characteristics are important, with the $R^2$ increasing from roughly 0.10 to nearly 0.50 as more respondent history is provided.

Third, the design generalizes beyond the focus of our main analysis of economy-wide optimism. For the respondent's optimism about their own firm, the LLM score performs similarly and leaves no incremental role for the respondent's own prior answer. For the more volatile expected revenue growth, it remains a positive predictor, although weaker statistically.

Fourth, the relation between human CFO sentiment and digital twin sentiment is robust to aggregation. At the quarterly level, all three LLM measures significantly predict aggregated CFO expectations.

*Limitations and caveats.* There are several limitations to our analysis. First, even in the specifications with the best fit, there is substantial unexplained variation in CFO expectations. Actual CFOs possess private information, such as pending orders, internal cost pressures, and board-level strategy shifts, that a model conditioned on public data and survey history can not fully recover. Thus, the LLM complements the human survey rather than replaces it (Wang et al., 2026; Dillion et al., 2023). Second, the digital twin performs best when it is conditioned on a history of human answers to previous surveys. Even if one were satisfied with the LLM's performance to date, if the human CFOs were to cease producing survey expectations, the ability of LLMs to mimic CFO expectations likely would deteriorate without new human data going forward. That is, the LLM relies on humans to produce future expectations on which it can condition. Third and related to this, the weak predictive ability of the LLM for first-time CFOs means that attempting to expand the effort to include a large number of new companies (with no historical survey information to condition on) may produce unreliable results, at least with current LLM models.

Fourth, while we show that twins match their human counterparts' answers, the process the LLM uses is a black box, and therefore we can not say whether the LLM mimics the thought process or behavioral traits of their twin CFO. Fifth, like any forecasting exercise, the model is a product of its training data, and its "beliefs" may embed historical regularities that break down in structurally novel regimes.

Finally, although our information-cutoff strategy and the fact that the target was never in the training corpus attenuate the look-ahead concern of Lopez-Lira et al. (2025), we cannot fully rule out indirect contamination. Two avenues would further alleviate this concern: training a time-stamped model at each point in time (Sarkar, 2024), which is computationally demanding (He et al., 2026); and out-of-corpus validation conducted live at the time of future quarterly surveys.

*Possible applications using digital twins.* If these limitations are adequately addressed, and digital twins can accurately mimic a wide-range of companies and settings, there are a number of potential applications for digital twins. Unlike a typical human survey that occurs at a fixed periodicity (e.g., quarterly), researchers, practitioners, and policymakers could conduct a synthetic survey very quickly and at any point in time. Further, one could perform counterfactual experiments that explore changes in macro or firm-specific conditions (e.g., recession; varying foreign-revenue exposure while holding everything else fixed). Policy-makers could thus conduct a wide range of stress tests - perhaps even creating a synthetic Beige Book.

## References

Aher, G., R. I. Arriaga, and A. T. Kalai (2023). Using large language models to simulate multiple humans and replicate human subject studies. In *Proceedings of the 40th International Conference on Machine Learning*, Volume 202 of *PMLR*, pp. 337–371. https://doi.org/10.48550/arXiv.2208.10264.

Argyle, L. P., E. C. Busby, N. Fulda, J. R. Gubler, C. Rytting, and D. Wingate (2023). Out of one, many: Using language models to simulate human samples. *Political Analysis 31*(3), 337–351. https://doi.org/10.1017/pan.2023.2.

Barrero, J. M. (2022). The micro and macro of managerial beliefs. *Journal of Financial Economics 143*(2), 640–667. https://doi.org/10.1016/j.jfineco.2021.06.007.

Baslandze, S., Z. Edwards, J. R. Graham, T. McClure, B. H. Meyer, M. Sparks, S. R. Waddell, and D. Weitz (2026). Artificial intelligence, productivity, and the workforce: Evidence from corporate executives. NBER Working Paper 34984. Forthcoming, *Journal of Finance: Insights and Perspectives*. https://doi.org/10.3386/w34984.

Ben-David, I., J. R. Graham, and C. R. Harvey (2013). Managerial miscalibration. *Quarterly Journal of Economics 128*, 1547–1584. https://doi.org/10.1093/qje/qjt023.

Bloom, N. and J. Van Reenen (2007). Measuring and explaining management practices across firms and countries. *Quarterly Journal of Economics 122*(4), 1351–1408. https://doi.org/10.1162/qjec.2007.122.4.1351.

Boutros, M., I. Ben-David, J. R. Graham, C. R. Harvey, and J. W. Payne (2026). The persistence of miscalibration. *Review of Financial Studies 39*(11), 3413–3449. https://doi.org/10.1093/rfs/hhaf070.

Brown, W. (1910). Some experimental results in the correlation of mental abilities. *British Journal of Psychology 3*(3), 296–322. https://doi.org/10.1111/j.2044-8295.1910.tb00207.x.

Cameron, A. C., J. B. Gelbach, and D. L. Miller (2011). Robust inference with multiway clustering. *Journal of Business & Economic Statistics 29*(2), 238–249. https://doi.org/10.1198/jbes.2010.07136.

Chetty, R., J. Fogel, L. F. Katz, K. Noray, S. Porter, and J. Reisinger (2026). Experimentally validated digital twins: Evaluating workforce training programs in the age of big data. https://arxiv.org/abs/2611.03938.

Coibion, O., Y. Gorodnichenko, and S. Kumar (2018). How do firms form their expectations? New survey evidence. *American Economic Review 108*(9), 2671–2713. https://doi.org/10.1257/aer.20151299.

Cook, T. R., S. Kazinnik, A. L. Hansen, and P. McAdam (2023). Evaluating local language models: An application to bank earnings calls. Research Working Paper 23-12, Federal Reserve Bank of Kansas City. https://doi.org/10.18651/RWP2023-12.

Dillion, D., N. Tandon, Y. Gu, and K. Gray (2023). Can AI language models replace human participants? *Trends in Cognitive Sciences 27*(7), 597–600. https://doi.org/10.1016/j.tics.2023.04.008.

Fried, S., T. Graf, S. Malhotra, and S. R. Singh (2026, May). What do financial officers predict for price growth? *FRBSF Economic Letter 2026*(11). Federal Reserve Bank of San Francisco. https://www.frbsf.org/research-and-insights/publications/economic-letter/2026/05/what-do-financial-officers-predict-for-price-growth/.

Fritsch, G. P. and J. Z. Mazlish (2026). High-frequency fiscal shocks. Working paper, University of Oxford. https://jzacharymazlish.com/files/High_frequency_fiscal_shocks.pdf.

Gao, Z., W. Jiang, and Y. Yan (2026). Detecting lookahead bias in LLM forecasts. Working Paper, SSRN 5985277. https://doi.org/10.2139/ssrn.5985277.

Gennaioli, N., Y. Ma, and A. Shleifer (2016). Expectations and investment. *NBER Macroeconomics Annual 30*, 379–431. https://doi.org/10.1086/685965.

Goulding, C. L., C. R. Harvey, and H. Kurtović (2026a). Disagreement of disagreement. *Journal of Financial Economics 185*, 104379. https://doi.org/10.1016/j.jfineco.2026.104379.

Goulding, C. L., C. R. Harvey, and H. Kurtović (2026b). The price of conviction. *SSRN Working Paper 7571098*. https://doi.org/10.2139/ssrn.7571098.

Graham, J., B. Meyer, N. Parker, and S. R. Waddell (2020, May). Introducing The CFO Survey: A collaboration between Duke University's Fuqua School of Business and the Federal Reserve Banks of Richmond and Atlanta. https://www.richmondfed.org/research/national_economy/cfo_survey/research_and_commentary/2020/20200515_introducing_cfo_survey. Accessed: April 3, 2026.

Graham, J. R. (2022). Presidential address: Corporate finance and reality. *Journal of Finance 77*, 1975–2049. https://doi.org/10.1111/jofi.13161.

Graham, J. R. and C. R. Harvey (2001). The theory and practice of corporate finance: Evidence from the field. *Journal of Financial Economics 60*, 187–243. https://doi.org/10.1016/S0304-405X(01)00044-7.

Hansen, A. L., J. J. Horton, S. Kazinnik, D. Puzzello, and A. Zarifhonarvar (2025). Simulating the survey of professional forecasters. Working Paper, SSRN 5066286. https://doi.org/10.2139/ssrn.5066286.

He, S., L. Lv, A. Manela, and J. Wu (2026). Chronologically consistent large language models. *Journal of Financial Economics 185*, 104374. https://doi.org/10.1016/j.jfineco.2026.104374.

Horton, J. J., A. Filippas, and B. S. Manning (2023). Large language models as simulated economic agents: What can we learn from Homo Silicus? NBER Working Paper 31122. https://doi.org/10.3386/w31122.

Li, P., N. Castelo, Z. Katona, and M. Sarvary (2024). Frontiers: Determining the validity of large language models for automated perceptual analysis. *Marketing Science 43*(2), 254–266. https://doi.org/10.1287/mksc.2023.0454.

Lopez-Lira, A., Y. Tang, and M. Zhu (2025). The memorization problem: Can we trust LLMs' economic forecasts? Working Paper, SSRN 5217505. https://doi.org/10.2139/ssrn.5217505.

Ludvigson, S. C. (2004). Consumer confidence and consumer spending. *Journal of Economic Perspectives 18*(2), 29–50. https://doi.org/10.1257/0895330041371222.

Ludwig, J., S. Mullainathan, and A. Rambachan (2026). Large language models: An applied econometric framework. *Annual Review of Economics 18*, 283–316. https://doi.org/10.1146/annurev-economics-120925-105620.

Malmendier, U. and G. Tate (2015). Behavioral CEOs: The role of managerial overconfidence. *Journal of Economic Perspectives 29*, 37–60. https://doi.org/10.1257/jep.29.4.37.

Manning, B. S. and J. J. Horton (2026). General social agents. NBER Working Paper 34937. https://doi.org/10.3386/w34937.

Manning, B. S., K. Zhu, and J. J. Horton (2024). Automated social science: Language models as scientist and subjects. NBER Working Paper 32381. https://doi.org/10.3386/w32381.

Mei, Q., Y. Xie, W. Yuan, and M. O. Jackson (2024). A Turing test of whether AI chatbots are behaviorally similar to humans. *Proceedings of the National Academy of Sciences 121*(9), e2313925121. https://doi.org/10.1073/pnas.2313925121.

Newey, W. K. and K. D. West (1987). A simple, positive semi-definite, heteroskedasticity and autocorrelation consistent covariance matrix. *Econometrica 55*(3), 703–708. https://doi.org/10.2307/1913610.

Park, J. S., C. Q. Zou, J. Kamphorst, N. Egan, A. Shaw, B. M. Hill, C. Cai, M. R. Morris, P. Liang, R. Willer, and M. S. Bernstein (2026). LLM agents grounded in self-reports enable general-purpose simulation of individuals. Working Paper, arXiv:2411.10109. https://doi.org/10.48550/arXiv.2411.10109.

Santurkar, S., E. Durmus, F. Ladhak, C. Lee, P. Liang, and T. Hashimoto (2023). Whose opinions do language models reflect? In *Proceedings of the 40th International Conference on Machine Learning*, Volume 202 of *PMLR*, pp. 29971–30004. https://doi.org/10.48550/arXiv.2303.17548.

Sarkar, S. K. (2024). StoriesLM: A family of language models with time-indexed training data. Working Paper, SSRN 4881024. https://doi.org/10.2139/ssrn.4881024.

Shrout, P. E. and J. L. Fleiss (1979). Intraclass correlations: Uses in assessing rater reliability. *Psychological Bulletin 86*(2), 420–428. https://doi.org/10.1037/0033-2909.86.2.420.

Spearman, C. (1910). Correlation calculated from faulty data. *British Journal of Psychology 3*(3), 271–295. https://doi.org/10.1111/j.2044-8295.1910.tb00206.x.

Suh, J., E. Jahanparast, S. Moon, M. Kang, and S. Chang (2025). Language model fine-tuning on scaled survey data for predicting distributions of public opinions. In *Proceedings of the 63rd Annual Meeting of the Association for Computational Linguistics (Volume 1: Long Papers)*, Vienna, Austria, pp. 21147–21170. Association for Computational Linguistics. https://doi.org/10.18653/v1/2025.acl-long.1028.

Vafa, K., A. Rambachan, and S. Mullainathan (2024). Do large language models perform the way people expect? Measuring the human generalization function. In *Proceedings of the 41st International Conference on Machine Learning*, Volume 235 of *PMLR*, pp. 48919–48937. https://doi.org/10.48550/arXiv.2406.01382.

Wang, M., D. J. Zhang, and H. Zhang (2026). Large language models for market research: A data-augmentation approach. *Marketing Science 45*(4), 728–751. https://doi.org/10.1287/mksc.2025.0009.

Wu, J. C., J. Xi, and S. Xie (2025). LLM survey framework: Coverage, reasoning, dynamics, identification. NBER Working Paper 34308. https://doi.org/10.3386/w34308.

# Internet Appendix

CFOs Meet LLMs

## Internet Appendix A. Data Construction

This appendix describes the procedure for constructing the sample used in the estimation of the paper, beginning with the raw survey responses from the CFO survey. The CFO Survey is a survey of corporate financial officers conducted by Duke University prior to June 2020 and jointly by Duke and the Federal Reserve Banks of Atlanta and Richmond since then.

### *A.1 Sample Selection*

We start with the universe of survey responses from 2002 to 2025. We require that the respondent works for a publicly traded firm, so that the LLM can find public information on the firm, such as filings, earnings calls, and analyst coverage. We also require that the respondent answers the question about economic optimism. Using these requirements, we match a total of 6,075 firm-quarter observations for our estimation.

### *A.2 Company Name Standardization*

Since we repeatedly survey the same population, the same company can be mentioned under different variations of its name, e.g., "Acade Corp.", "ACADE CORPORATION" and "Acade, Inc.". To track companies across the survey, we apply the following procedure to standardize company names:

1. Convert all characters to uppercase.
2. Remove common legal suffixes, including Inc., Corp., Corporation, LLC, LLP, Ltd., L.P., Co., and their variations (with or without trailing periods).
3. Eliminate parenthetical text (e.g., "(US)" or "(formerly XYZ)").
4. Strip leading and trailing whitespace, collapse consecutive internal spaces to a single space, and remove extraneous punctuation such as trailing commas or periods.

After standardization, we assign a unique firm identifier (PERMCO, a permanent company identifier from the Center for Research in Security Prices) to each distinct cleaned name. This identifier is used to construct the firm panel and to apply firm fixed effects in Table 2.

### *A.3 Respondent Linkage and History*

To create each respondent's history, we first create a person key based on the available identifiers in the sample (either the survey's internal respondent identifier or else a standardization of the contact field). We then fill identity and profile fields for each person key. For each respondent and survey date, we keep that respondent's responses dated strictly before the survey date and within the preceding 12 calendar quarters. From these we supply (i) up to 12 of the respondent's past optimism scores; (ii) the LLM's earlier forecasts for the same respondent in the same quarters, each the mean of three calls; and (iii) the respondent's average optimism over those quarters. The runner processes calendar quarters in order, so that the LLM's predictions for earlier quarters of a respondent's life are available for predicting later quarters.

### *A.4 Missing Data Imputation*

The survey instrument collects several firm-level characteristics at each wave: job title, industry sector, sales revenue, total employment, percentage of foreign sales, and headquarters location. However, not all respondents complete every field in every wave.

To maximize the information we provide to the LLM, we impute missing values using the temporally closest non-missing value for each firm. We take this approach because we believe that firm characteristics such as industry, location, revenue, and number of employees change slowly over time, and the passage of a single quarter is unlikely to induce a change in these broad categories. If a missing value persists after imputation, we provide the LLM with whatever information is available.

### *A.5 Variable Definitions*

Table IA1 summarizes the key variables used in the analysis.

**Table IA1:** Variable Definitions

| **Variable** | **Definition** |
|---|---|
| CFO Survey Optimism | Self-reported optimism about the overall U.S. economy on a 0–100 scale, from the Duke–Federal Reserve CFO Survey. |
| LLM Optimism | A numeric value from 0 to 100 that reflects how optimistic GPT (model alias `gpt-5.4`, run on April 21, 2026) would be about the overall US economy if the model were the CFO of the given firm on the survey date. This score is based on the firm's profile and the respondent's past survey responses before the cutoff date, and represents the average from three replications. |
| Last-Quarter CFO Optimism | The same respondent's CFO Survey optimism score from the immediately preceding survey wave (used in Table 2, Panel A, columns (5)–(8)). The any-prior-quarter analog is used in Table IA3. |
| Own-Firm Optimism | Self-reported optimism about the financial prospects of the respondent's own company, on a 0–100 scale, from the Duke–Federal Reserve CFO Survey. The matched LLM analog is generated with the same prompt design, substituting this question and the respondent's history of answers to it (Table 3, Panel A). |
| Expected Revenue Growth | Respondent's expected growth rate of their own firm's revenue over the next twelve months, in percent. The matched LLM analog is constructed as above (Table 3, Panel B). |
| Michigan Consumer Sentiment | University of Michigan Index of Consumer Sentiment (`ics_all`), quarterly. |
| SPF Real GDP Growth | Mean forecast of year-ahead real GDP growth (`drgdp5`) from the Survey of Professional Forecasters, Federal Reserve Bank of Philadelphia. |

### *A.6 Quarterly Aggregation*

In Table 4 we aggregate and analyze the data at the quarterly level. For each year-quarter we take the average of the LLM optimism scores and of the CFO Survey optimism scores, and we match them with the contemporaneous values of the Michigan Consumer Sentiment Index and the year-ahead real GDP growth rate forecast from the Survey of Professional Forecasters consensus. This gives us 93 observations.

## Internet Appendix B. System Prompt

Financial executives provide forecasts about their expectations of the future, from the perspective of their own companies, in The CFO Survey, which is jointly conducted by Duke University and the Federal Reserve Banks of Richmond and Atlanta.

Your task is to prepare forecasts as if you are a financial executive working at a specific company at a specific point in time, and then repeat the task as if you are a different financial executive working at a different company, and keep repeating this task the specified number of times.

When preparing forecasts: (i) use only the information available as of the `CUTOFF_DATE`; (ii) search for information about the specific company you are role-playing—if detailed information is not available, use information on comparable companies, drawing from financial statements, analyst reports, and industry data; (iii) if the profile includes this person's own past survey responses, treat them as a strong prior on this individual's style, optimism bias, and level—your answer should be consistent with their personal pattern unless clearly contradicted by new information as of the `CUTOFF_DATE`; (iv) if the profile includes the prior LLM forecasts you have previously produced for this same person, treat those as a secondary prior—continuity matters, but the actual human responses (when present) take precedence; (v) explicitly incorporate the following dimensions into your thought process.

*Historical Financial Performance:* review past 3–5 years of revenue data (quarterly and annual); identify growth trends, seasonality patterns, and cyclical behaviors; analyze revenue composition by business segments, product lines, or geographic regions; examine profit margins and their relationship to revenue growth.

*Forward-Looking Company Indicators:* latest management guidance and earnings call transcripts; product pipeline, launch schedules, and innovation roadmap; customer acquisition trends, retention rates, and churn analysis; order backlog, booking trends, and contract renewals; expansion plans; sales force changes and marketing investment levels.

*Analyst Coverage:* collect and synthesize recent analyst reports and price targets; review consensus revenue estimates and any recent revisions; identify key assumptions driving analyst projections; note any divergent views or concerns raised by analysts.

*Comparable Company Analysis (if primary data unavailable):* focus on industry peers with similar business models, revenue streams, market capitalization, geographic footprint, and customer base; analyze revenue growth patterns, market share trends, and competitive posi-tioning; evaluate responses to industry headwinds and tailwinds; review valuation multiples, growth assumptions, and recent strategic initiatives.

*Market and Industry Context:* overall industry growth projections and market size; key industry trends, disruptions, and regulatory changes; competitive landscape and market share dynamics; customer demand patterns and buyer behavior shifts; technology adoption rates and innovation cycles.

*Macroeconomic Factors:* GDP growth forecasts for relevant geographic markets; interest rate environment and credit conditions; currency exchange rate impacts (for international operations); inflation effects on pricing power and costs; sector-specific economic indicators.

For each simulation, assume the role of one survey participant and provide forecasts. Incorporate all relevant details about overall economic conditions up to the `CUTOFF_DATE` and exclude any data or developments occurring after the `CUTOFF_DATE`. Internally consider historical performance, company guidance, analyst views, industry conditions, macro factors, and the respondent's own history—think step-by-step before answering. Do *not* output or reveal internal chain-of-thought or reasoning. For each simulated CFO produce exactly one line of output in this exact format: `<number>`, where `<number>` is the forecast number only (no percent sign, no decimals unless necessary).

*Survey Question:* Rate your optimism about the overall U.S. economy on a scale from 0–100, with 0 being the least optimistic and 100 being the most optimistic.

## Internet Appendix C. Sample Query

`CUTOFF_DATE` for this exercise: 2024-02-15

Profile for this exercise:

Name: [confidential]
Current role: CFO
Company: [confidential]
Ticker: [confidential]
Industry: [confidential]
Sales revenue: $1–4.9 billion
Number of employees: 2,500–4,999
Percent of sales in foreign countries: More than 50%

Headquarters location: Midwest U.S.
State: [confidential]
Ownership: Public
Credit rating: [confidential]

Respondent history:

This person's own past U.S.-economy optimism responses (0–100), up to 12 most recent prior responses, within the last 12 calendar quarters:

- 2023Q1: 55.0
- 2023Q2: 58.0
- 2023Q4: 60.0

Your (the LLM's) prior forecasts for this same person (mean of 3 runs), up to 12 most recent prior responses, within the last 12 calendar quarters:

- 2023Q1: 54.3
- 2023Q2: 56.7
- 2023Q4: 59.0

This person's average U.S.-optimism response across the last 12 calendar quarters (N=3): 57.7

## Internet Appendix D. Forecast Stability and Reproducibility

As discussed in Section I, we run the LLM three times with the same prompt and average the calls. Table IA2 reports the standard deviation of LLM calls to the same prompt. On average, the standard deviation is 2.22, which is considerably smaller than the average level of optimism in the LLM forecast of 56.6. Averaging three independent noisy measurements reduces the noise and the test-retest reliability of the average score that we use throughout this paper is 0.985. In other words, only 1.5 percent of the variance of the aggregate score is due to random fluctuations across LLM calls.

Note that the numbers reported in Table IA2 represent intra-run variations. Running the same prompts at a later date will likely result in different numbers, since `gpt-5.4` is merely a vendor-supplied alias, and the actual version of GPT that the vendor deploys can change over time without notice. We report the model alias as well as the date on which we started the runs (April 21, 2026). Our open-weight replication in Internet Appendix F does not suffer from this issue since the Gemma weights are fixed.

**Table IA2:** Stability of LLM-Generated Forecasts Across Replications

This table reports within-prompt stability statistics for the LLM-generated optimism scores. For each of the $N$ =6,075 survey observations, three independent forecasts are generated by querying the LLM through its API with identical prompts. Pairwise correlations are computed across all replication pairs; the within-prompt standard deviation and coefficient of variation (CV) are computed row-wise and then averaged over all survey observations. The intraclass correlation coefficient (ICC) is estimated via one-way ANOVA (Shrout and Fleiss, 1979) and measures the reliability of a single API call. The reliability of an average of $n$ calls follows from the Spearman–Brown formula (Spearman, 1910; Brown, 1910), $\text{ICC}_n = n\,\text{ICC}/[1+(n-1)\text{ICC}]$.

| **Statistic** | **Value** |
|---|---|
| *Pairwise Correlations Across Replications* | |
| Replication 1 vs. Replication 2 | 0.958 |
| Replication 1 vs. Replication 3 | 0.959 |
| Replication 2 vs. Replication 3 | 0.957 |
| *Within-Prompt Variation* | |
| Mean standard deviation | 2.22 |
| Mean coefficient of variation | 4.9% |
| *Intraclass Correlation (One-Way ANOVA)* | |
| ICC | 0.958 |
| 95% confidence interval | [0.956, 0.959] |
| Reliability of group mean ($n \approx 3$) | 0.985 |

## Internet Appendix E. Controlling for Any Prior Response

Table 2 (columns (5)–(8)) controls for the respondent's answer in the *immediately preceding* survey wave. Table IA3, on the other hand, reports the same specifications but instead controls for the most recent prior optimism from any prior quarter for which that is available (i.e. the respondent's prior optimism on record). This allows for a larger sample to be used. Even in these specifications, the LLM score is a strong predictor.

**Table IA3:** Predicting CFO Optimism, Controlling for Any Prior Response

This table contains results from columns (5)–(8) of Table 2 when the sample is restricted to those who have at least one prior response to the survey, and the estimation includes a control for the individual's most recent prior optimism recorded by the survey (*Prior CFO Optimism*), which may not be from the prior survey. The dependent variable is an individual's optimism from the CFO Survey in each column. Columns include additional firm fixed effects and year-quarter fixed effects. The number of observations decreases in the columns that include firm fixed effects as firms who only appear in one survey are dropped. Two-way cluster-robust standard errors (firm and year-quarter) are reported with the corresponding $t$-statistics in parentheses. *, **, and *** indicate significance at the 10%, 5%, and 1% levels, respectively.

| | *Dep. variable = CFO Survey Optimism* | | | |
|---|---|---|---|---|
| | (1) | (2) | (3) | (4) |
| LLM Optimism | 0.583*** | 0.564*** | 0.507*** | 0.312*** |
| | (12.92) | (12.56) | (7.60) | (4.32) |
| Prior CFO | 0.0880** | −0.0828** | 0.140*** | 0.0483 |
| Optimism | (2.11) | (−2.07) | (2.64) | (1.09) |
| Firm FE | | Y | | Y |
| Year-Quarter FE | | | Y | Y |
| $N$ | 3,832 | 3,624 | 3,832 | 3,624 |
| $R^2$ | 0.397 | 0.519 | 0.434 | 0.566 |

## Internet Appendix F. Replication with an Open-Weight Model

The results in the main text are based on a single LLM, a proprietary frontier model that we access by API. Here, we re-run the entire exercise with an open-weight LLM that can be hosted locally. We use Gemma 4, developed by Google DeepMind, and run it on the high-performance computing cluster at Georgia State University (the specific model is gemma-4-31B). We prompt Gemma in the exact same way as described in Section I, with the same information cutoffs and conditioning on each respondent's response history. We obtain responses for the same 6,075 matched responses used in the main analysis. Because Gemma is hosted locally, it does not have access to a retrieval tool and it relies solely on its internal knowledge from its training corpus, the firm profile provided in the prompt, and the respondent's response history. This allows us to check not only if the results are sensitive to the choice of LLM, but also to conduct a test where the LLM is not allowed to use web search. Prior research has shown that locally-hosted, open-source LLMs can perform on par with cloud-hosted, proprietary LLMs on text-based tasks in finance (Cook et al., 2023).

Table IA4 replicates the specifications of Table 2, replacing the GPT-5.4 score with the Gemma 4 score. The results are very similar across the two LLMs. For instance, in the baseline specification without controls, the coefficient on the LLM score is 0.572 for Gemma 4, compared to 0.565 for GPT-5.4 (with $R^2$ of 0.263 for Gemma 4 versus 0.269 for GPT-5.4). In the last specification, which includes firm and year-quarter fixed effects and the respondent's last answer, the coefficient on the LLM score is 0.282 for Gemma 4, compared to 0.286 for GPT-5.4. The main difference we observe is that the last answer by the same respondent becomes insignificant in three of the four specifications where it is included. We read the table as evidence that the individual-level predictive content is a property of the conditioning design rather than of any one model or vendor.

**Table IA4:** Predicting Individual CFO Survey Optimism: Open-Weight Model

This table is identical to Table 2, Panel A except that Gemma 4 (gemma-4-31B), rather than GPT-5.4, is used to generate the optimism scores. Gemma 4 is an open-weight model that was run on a local server for this analysis. The prompts fed to Gemma 4, the information cutoff date, conditioning on the individual's past responses, and the estimation sample are identical to the main table. Gemma 4 does not use the web search tool. The dependent variable is the individual CFO Survey respondent's optimism score, measured on a scale of 0 to 100. Columns 1 to 4 use the full sample of matched observations. Columns 5 to 8 restrict the sample to CFOs who were also surveyed in the immediately previous survey and controlled for their response to that survey, *Last-Quarter CFO Optimism*. The number of observations drops in the firm fixed-effect columns because firms that are only ever surveyed one time are dropped when their observation is absorbed by their fixed effect. Standard errors are clustered by both firm and year-quarter. $t$-statistics are in parentheses. *, **, and *** indicate statistical significance at the 10%, 5%, and 1% levels, respectively.

| | *Dep. variable = CFO Survey Optimism* | | | | | | | |
|---|---|---|---|---|---|---|---|---|
| | No History Control | | | | Control for Last-Quarter Resp. | | | |
| | (1) | (2) | (3) | (4) | (5) | (6) | (7) | (8) |
| LLM Optimism (Gemma) | 0.572*** | 0.411*** | 0.550*** | 0.266*** | 0.579*** | 0.617*** | 0.435*** | 0.282* |
| | (21.72) | (11.68) | (20.08) | (7.54) | (8.22) | (6.97) | (3.00) | (1.91) |
| Last-Quarter | | | | | 0.110 | −0.119 | 0.247* | 0.129 |
| CFO Optimism | | | | | (1.35) | (−1.18) | (1.70) | (0.91) |
| Firm FE | | Y | | Y | | Y | | Y |
| Year-Quarter FE | | | Y | Y | | | Y | Y |
| $N$ | 6,075 | 5,623 | 6,075 | 5,623 | 1,982 | 1,820 | 1,982 | 1,820 |
| $R^2$ | 0.263 | 0.436 | 0.307 | 0.495 | 0.486 | 0.601 | 0.539 | 0.660 |